\documentclass[preprint,12pt]{elsarticle}

\usepackage[T1]{fontenc}

\usepackage{graphicx} 
\usepackage{amssymb}
\usepackage{amsmath}
\usepackage{subfig}
\usepackage{tikz}
\usepackage{transparent}

\journal{}

\begin{document}

\begin{frontmatter}

\title{3D morphology formation in a mixture of three differently
averse components
}
\author[EC]{Emilio N.M. Cirillo}
\affiliation[EC]{organization={Dipartimento di Scienze di Base e Applicate per l'Ingegneria, Sapienza Universit\`a di Roma},
            addressline={via Antonio Scarpa 16}, 
            city={Roma},
            postcode={00161}, 
            country={Italia}}
\author[NJ]{Nicklas J\"averg\r{a}rd}
\affiliation[NJ]{organization={Department of Mathematics and Computer Science, Karlstad University},
            city={Karlstad},
            country={Sweden}}  
  \author[RL]{Rainey Lyons}
\affiliation[RL]{organization={Department of Applied Mathematics, University of Colorado Boulder},
            city={Boulder},
            country={USA}}  
            
\author[NJ]{Adrian Muntean}
            
\author[SAM]{Stela Andrea  Muntean}
\affiliation[SAM]{organization={Department of Engineering and Physics, Karlstad University},
            city={Karlstad},
            country={Sweden}}  

\begin{abstract}
Film formation from solvent evaporation in polymer ternary solutions
is relevant for several technological applications,
such as the fabrication of organic solar cells.
The performance of the final device will strongly depend on the internal morphology of the obtained film, which, in turn, is affected by the processing conditions.
We are interested in modeling morphology formation in 3D for ternary mixtures using both a lattice model
and its continuous counterpart in the absence of evaporation.
In our previous works,  we found that, in 2D, both models
predict the existence of two distinct regimes: (i)
a low-solvent regime, characterized by two interpenetrated domains
of the two polymers, and (ii) a high-solvent regime, where isolated
polymer domains are dispersed in the solvent background.
In the significantly more intriguing 3D case, we observe a
comparable scenario both for the discrete and the 
continuous model. The lattice
model reveals its ability to describe morphology formation 
even in the high solvent content 3D case, in which the three-dimensional nature
of space could have prevented cluster formation.
\end{abstract}



\end{frontmatter}

\date{March 2025}

\section{Introduction}
Morphology formation in solution processed thin films is crucial for organic solar cells (OSCs) as it strongly affects their efficiency and stability\cite{hoppe2004organic,Moons2002,Moons2007}. 
A well-optimized nanoscale morphology enables efficient exciton separation and charge movement in organic photovoltaic devices while minimizing losses caused by recombination.  
These morphologies can also influence processability, scalability, and light absorption, hence further impacting overall device performance. Efforts are made to follow the {\em in situ} evolution of the morphology \cite{Nannan2023_InSitu}, modeling and simulations playing a crucial role in completing the insights in this direction \cite{MichelsMoons2013,VikasWodo2018,Harting_ATS}. However, in spite of the enormous volume of literature concerning the  phase separation of binary mixtures in 2D, achieving a good theoretical understanding of morphology formation in multi-component systems in 3D remains a significant challenge.

We aim to generate various morphology classes suitable for the study of OSCs through numerical simulations. Notably, our findings may also be relevant to other composite thin-film materials, such as adhesive bands \cite{Kronberg_2023}.
While some partial results exist for Cahn-Hilliard-type systems, general conclusions have yet to be established (see, e.g., \cite{giacomin1998phase,Voorhees}). Numerical simulations, however, offer valuable insights into specific scenarios. 
For related studies on the numerical simulation of the Cahn-Hilliard equation in 3D, we refer the reader to \cite{bailo2023,Zhou2021, Xiao_2023,NGLBS2024}.

In this work, we present our latest findings on the numerical simulation of morphology formation in ternary systems in 3D, where one component acts as a background solvent, effectively shielding the other two, which strongly repel each other. 
The reference case we consider cannot be directly captured by straightforward generalizations of the Cahn-Hilliard theory for three-component phase separation. 
Instead, we have previously demonstrated that, at least in 2D, the Blume-Capel particle model with Kawasaki dynamics (compare \cite{Andrea_EPJ, Andrea_PhysRevE,Euromech}), along with a continuum counterpart—a system of two strongly coupled, nonlocal, nonlinear drift-diffusion equations (\cite{Marra,lyons2024phase,lyons2023continuum})—provide an adequate description of this phenomenon. 
These models, therefore, serve as useful tools for exploring phase separation in such systems.

We efficiently succeeded to simulate both the discrete and the 
continuous model in 3D, performing a thorough qualitative analysis 
of their ability to produce morphologies. In this challenging 3D case, 
we observe a comparable scenario both for the discrete and 
continuous case, with respect to our previous 2D results \cite{lyons2023continuum,lyons2024phase}, namely the existence of a low--solvent regime in which the two polymers form interpenetrated domains, and a high--solvent regime where isolated polymer domains float in a solvent background.

When the models are run with 
high solvent content, the three-dimensional nature of space plays a 
key role in regulating cluster formation, since particles move following an
essentialy random motion. We show that even in this regime 
the 3D space structure does not 
prevent cluster formation.
These effects are crucial for understanding real--world applications, 
where film formation occurs in three-dimensional environments, and 
can provide deeper insights into optimizing processing conditions 
for polymer--based technologies.

This work is organized as follows: In Section  \ref{lattice_model} we present our stochastic lattice model as well as the 3D morphologies we obtain using it  for a selection of model parameters. In Section \ref{continuum_model}, we introduce the continuum model together with the corresponding 3D numerical simulations of the morphologies. Finally, we conclude this work in Section \ref{conclusion} with an outlook  on possible further developments around generating {\em in silico} 3D morphologies suitable for ensuring an optimized transport of charges and related applications.

\section{Morphology formation in 3D via a spin lattice model}\label{lattice_model}
To study morphology formation in a ternary mixture of two active components and one common solvent,
we adopt the Blume--Capel model, as in our previous papers \cite{Andrea_EPJ,Andrea_PhysRevE,Euromech}, where we investigated the 2D case.
As discussed above, in this paper we consider for the first time this 
problem in 3D. 
As much as possible, we shall adopt similar notation and use 
a similar language, in order to facilitate the reader willing to compare the results of this manuscript to the aforementioned previous work.
Moreover, in our graphical representations of the system configurations we shall use color codes analogous to the ones used in the previous papers. 

\subsection{Description of the lattice model}\label{lattice_model:model}
Let us
consider $\mathbb{Z}^3$ embedded in $\mathbb{R}^3$. 
We refer to its elements as \emph{sites} and, given two of them 
$i,i'\in\mathbb{Z}^3$, let $|i-i'|$ be their
Euclidean distance. 
Given $i\in\mathbb{Z}^3$, the site
$i'\in\mathbb{Z}^3$ is a \emph{nearest neighbor}
of $i$ if and only if $|i-i'|=1$. A \emph{bond} is a pair 
of neighboring sites.

We denote by $\Lambda$ the cubic torus 
$\{0,\dots,L-1\}^3\subset \mathbb{Z}^3$ and 
associate each site $i$ of $\Lambda$ with a spin variable $\sigma(i)$
taking values in the 
\emph{single spin state space} $\{-1,0,+1\}$.
The 
\emph{configuration} or \emph{state} space
is
$\mathcal{X}=\{-1,0,+1\}^{\Lambda}$.
The \emph{energy} of the configuration $\sigma\in\mathcal{X}$ is given 
by the Hamiltonian function $H:\mathcal{X}\to\mathbb{R}$ via
\begin{equation}
\label{eq:lat000}
H(\sigma)
=
J\sum_{\langle i,j\rangle}[\sigma(i)-\sigma(j)]^2
,
\end{equation}
with $J>0$,
where the sum is extended to the $3L^3$ bonds with periodic 
boundary conditions due to the torus topology.

According to the more mathematical oriented literature (see e.g., 
\cite{CO1996,cjs2024,cirillo2017sum}), 
the function \eqref{eq:lat000} is called the 
Hamiltonian of the Blume--Capel with zero magnetic field and chemical 
potential. 
We remark that \eqref{eq:lat000} would 
be addressed in the physics literature \cite{FGRN1994,blume1966}  as the 
Hamiltonian of the Blume--Capel model with 
ferromagnetic coupling $2J$, magnetic field zero, and crystal field $4J$.

In the usual spin model language, 
a site $i$ with $\sigma(i)=\pm1$ is considered occupied by a particle 
with spin $\pm1$, whereas it is considered empty if $\sigma(i)=0$.
In our polymer interpretation the $\pm1$ and $0$ spins 
will represent polymer and solvent molecules, respectively.

In Statistical Mechanics, the equilibrium properties 
of the system are described by the Gibbs 
measure 
$\exp\{-\beta H(\sigma)\}/\sum_{\eta\in\mathcal{X}}\exp\{-\beta H(\eta)\}$, where $\beta$ is a positive parameter which is interpreted as the \emph{inverse of the temperature}.
Its behavior has been widely studied and quite well understood in 2D, see, 
for instance, \cite{beale1986}. 
For the equilibrium properties of the 3D version of the model we refer
to the papers \cite{MFBprep2024,FT2013} and references therein.
The Blume--Capel model can be equipped with a dynamics, conservative (Kawasaki)
or not conservative (Glauber), reversible with respect 
to the Gibbs measure and several traditional questions,  
such metastability and spinodal decomposition, can be addressed. 
In 2D these problems have been quite widely studied, 
despite their intrinsic difficulty, due to the three--component character
of the model. We refer the reader to 
\cite{landim2016metastability,FGRN1994,MBR2024,cjs2024,cirillo2017sum,cirillo2013relaxation,CO1996,cjs2025}
for the metastable regime,  and respectively, to 
\cite{Euromech,Andrea_PhysRevE} for the spinodal regime. 

The reversibility assumption means that detailed 
balance is satisfied, namely, the product between the Gibbs factor 
of a configuration times the probability of 
a jump to a second configuration 
is equal to the Gibbs factor of the arrival configuration 
times the probablity of the backward transition. 
This assumption ensures that the dynamics has a stationary measure 
and that this measure is the Gibbs one, namely, the Gibbs measure 
describes the equilibrium of the stochastic model. 

Having in mind mixtures of two polymers and solvent, we choose 
the Blume--Capel model in view of the following peculiar property 
of its Hamiltonian: a direct 
interface between a minus and a plus spin costs $4J$, while 
an interface betwen a zero and a minus, or a plus, costs only $J$. 
On the other hand, bonds in which the two spins are alike
have zero energy contribution.
This makes the model a good choice to describe the morphological 
properties of a polymer blend in which two different types of
polymer, strongly repelling each other, are in the presence of 
a common solvent which tends to interpose among them acting as a shield. 

The Monte Carlo analysis that we propose here is purely qualitative 
and does not aim at studying the equilibrium and the non--equilibrium 
properties
of the model from the Statistical Mechanics perspective. Our sole
interest, for the moment, is to investigate the ability 
of the model to describe morphology formation in 3D starting from a 
completely random initial configuration. 
For this, we have to consider the model 
powered with a stochastic 
dynamics which preserves the value of the spins. 
A natural choice is, thus, the Gibbs measure reversible
Kawasaki dynamics:
bonds are selected at random, one at a time,
with uniform probability among those whose spins associated with
the sites defining the bond are different. Then,
these two spins 
are swapped with probability one if the variation of energy 
$\Delta$
due to the swap is non-positive or with probability 
$e^{-\beta \Delta}$ 
otherwise, that is to say, if $\Delta>0$.
We will call \emph{one iteration} of the dynamics the  
update of $3L^3$ bonds. 

Since the dynamics do not change the spin values, it is meaningful 
to let
$c_0$, $c_1$, and $c_{-1}=1-(c_0+c_1)$ be the fixed 
fraction of zeros, pluses, 
and minuses, respectively.
We remark again that, 
due to our interpretation of the system as a blend of polymers and solvent, 
we have been rather forced to use a conserved dynamics in which the spin values 
do not change during the evolution, but diffuse in the lattice 
thanks to the swapping mechanism. 

\begin{figure}[!t]
\begin{picture}(450,120)(-2,0)
\put(0,0)
{
 \includegraphics[width=.3\textwidth,height=.3\textwidth]{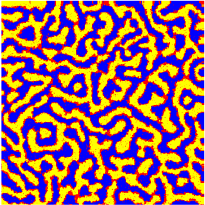}
}
\put(130,0)
{
 \includegraphics[width=.3\textwidth,height=.3\textwidth]{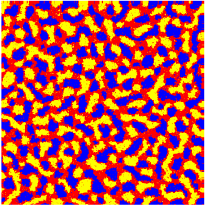}
}
\put(260,0)
{
 \includegraphics[width=.3\textwidth,height=.3\textwidth]{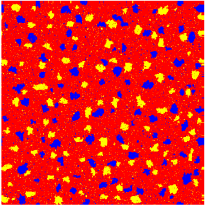}
}
\end{picture}
\caption{Configuration of the 2D version of the model on the $512\times512$ 
lattice for $J=1$ and $\beta=1.1$ after $10^5$ iterations. 
From the left to the right $c_0=0.2, 0.4, 0.8$.
In 2D, one iteration is the sequential update of $2L^2$ bonds.
Yellow, red, and blue points represent, respectively, minus, zero, and plus spins.
}
\label{fig:sto000}
\end{figure}

\subsection{Simulations}\label{lattice_model:simul}
We have demonstrated that this model is able 
to describe in 2D morphology formation and two different regimes have 
been identified. Indeed,  
starting from a completely random configuration in which spins are set 
equal to one, zero, or minus one with uniform probability, 
due to the structure of the Hamiltonian 
which favors configurations in which spins are  
surrounded by alike spins, 
provided the temperature is small enough (i.e., $\beta$ large enough), 
the dynamics evolve forming
domains of constant spin values. 
We found that shape and size of these regions  
strongly depend on the ratio of the spin mixture 
prescribed {\em a priori}. 

As illustrated in Fig.~\ref{fig:sto000}, at low solvent 
content ($c_0=0.2$ in the left picture) the system exhibits 
strongly oblonged minus and plus domains 
separated by a thin layer of zeroes showing a morphology 
reminiscent of the bicontinuous one typical of the two--state 
Ising model.
In contrast, 
at large solvent content ($c_0=0.8$ in the right picture), 
plus and minus clusters grow apart well separated by the 
background of zeros which fills the whole lattice. 
In the picture we have also reported the behavior of the 
intermediate $c_0=0.4$ case in which the morphology shares 
properties with both the extreme cases.
We refer to \cite{Andrea_PhysRevE,Euromech} for a detailed investigation
of these phenomena. 

\begin{figure}[!t]
\begin{picture}(450,400)(-2,0)
\put(0,0)
{
 \includegraphics[width=.3\textwidth,height=.3\textwidth]{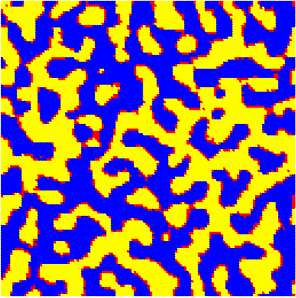}
}
\put(130,0)
{
 \includegraphics[width=.3\textwidth,height=.3\textwidth]{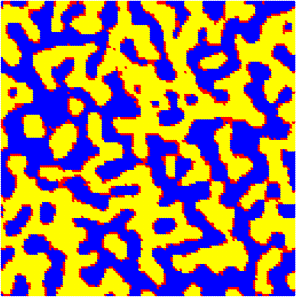}
}
\put(260,0)
{
 \includegraphics[width=.3\textwidth,height=.3\textwidth]{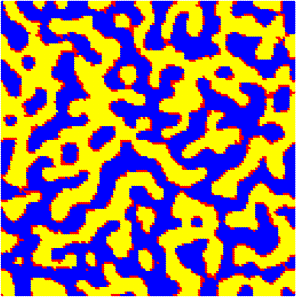}
}
\put(0,130)
{
 \includegraphics[width=.3\textwidth,height=.3\textwidth]{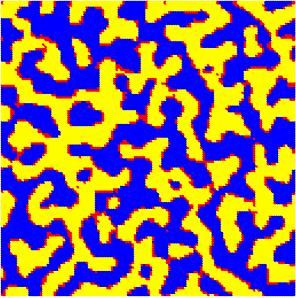}
}
\put(130,130)
{
 \includegraphics[width=.3\textwidth,height=.3\textwidth]{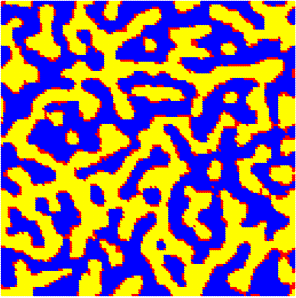}
}
\put(260,130)
{
 \includegraphics[width=.3\textwidth,height=.3\textwidth]{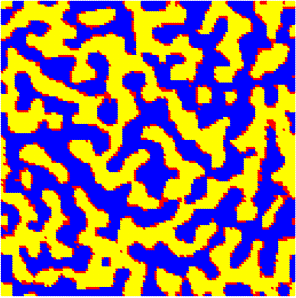}
}
\put(120,260)
{
 \includegraphics[width=.35\textwidth,height=.4\textwidth]{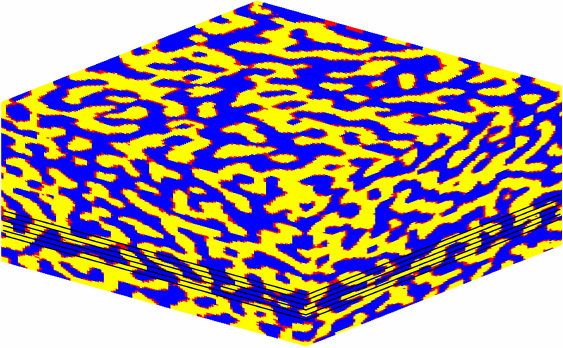}
}
\end{picture}
\caption{Configuration of the 3D Blume--Capel model 
on the $128^3$ 
lattice for 
$c_0=0.1$ after $4.8\times10^3$ iterations, with
$J=1$ and $\beta=1.0$. 
Top row: the left, right, and top planes are, respectively, the faces 
$(x,0,z)$, 
$(L-1,y,z)$, 
and $(x,y,L-1)$, 
with $x,y,z=0,\dots,L-1$, of the lattice $\Lambda$.  
Bottom rows: in lexicographical order sections at 
$z=20,24,28,32,36,40$ are shown (see, the black lines in the top panel).
Yellow, red, and blue points represent, respectively, minus, zero, and plus spins.}
\label{fig:sto010}
\end{figure}

\begin{figure}[!t]
\begin{picture}(450,400)(-2,0)
\put(0,0)
{
 \includegraphics[width=.3\textwidth,height=.3\textwidth]{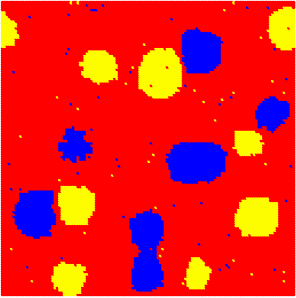}
}
\put(130,0)
{
 \includegraphics[width=.3\textwidth,height=.3\textwidth]{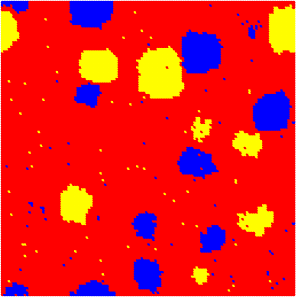}
}
\put(260,0)
{
 \includegraphics[width=.3\textwidth,height=.3\textwidth]{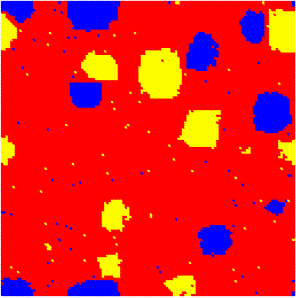}
}
\put(0,130)
{
 \includegraphics[width=.3\textwidth,height=.3\textwidth]{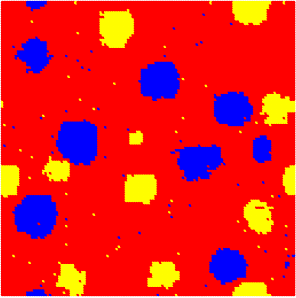}
}
\put(130,130)
{
 \includegraphics[width=.3\textwidth,height=.3\textwidth]{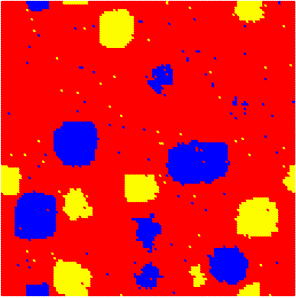}
}
\put(260,130)
{
 \includegraphics[width=.3\textwidth,height=.3\textwidth]{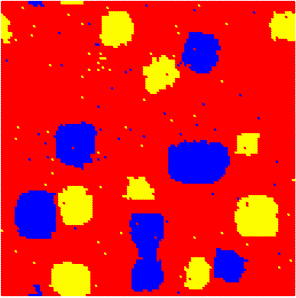}
}
\put(120,260)
{
 \includegraphics[width=.35\textwidth,height=.4\textwidth]{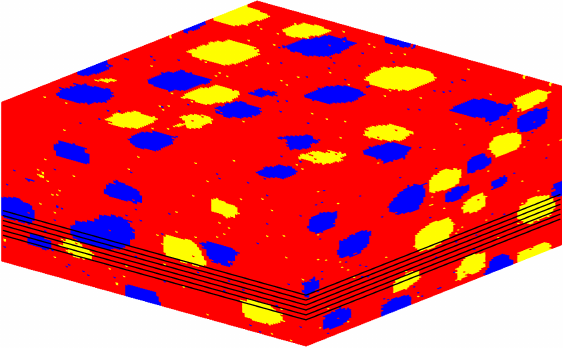}
}
\end{picture}
\caption{As in Fig.~\ref{fig:sto010} for
$c_0=0.8$.}
\label{fig:sto020}
\end{figure}

The 3D case is much more challenging, not only on the 
computational and visualization ground, as it is 
absolutely obvious, but 
also because we expect that the behavior of the dynamics 
will finely depend on the parameter of the model for large solvent 
content. 
Indeed, in such a case minuses and pluses move inside the zero background
performing grossly a symmetric random walk, since, being typically 
surrounded by zeroes, every attempted spins swap involves no energy change and, 
thus, it has probability one. Now, recalling that simple lattice 
random walks are 
recurrent in 2D and not recurrent in 3D, one easily understands that in our 
3D model, despite the fact that volume is finite, two pluses (resp.~minuses) have a very small 
probability to meet in a reasonable simulation time. 
Thus, we expect that in 3D, for large values of solvent content $c_0$,
the cluster formation is not a priori granted.

\begin{figure}[!t]
\begin{picture}(450,180)(-15,0)
\put(0,0)
{
 \includegraphics[width=.2\textwidth,height=.2\textwidth]{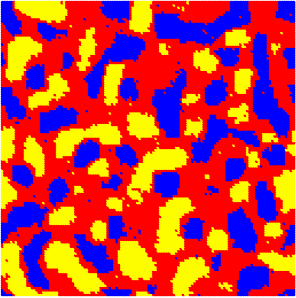}
}
\put(90,0)
{
 \includegraphics[width=.2\textwidth,height=.2\textwidth]{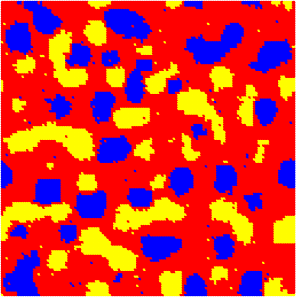}
}
\put(180,0)
{
 \includegraphics[width=.2\textwidth,height=.2\textwidth]{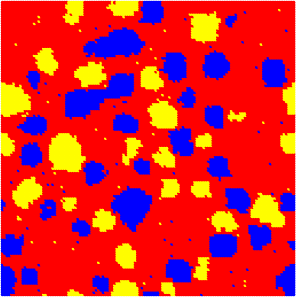}
}
\put(270,0)
{
 \includegraphics[width=.2\textwidth,height=.2\textwidth]{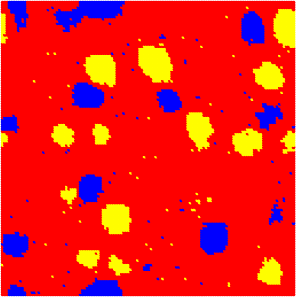}
}
\put(0,90)
{
 \includegraphics[width=.2\textwidth,height=.2\textwidth]{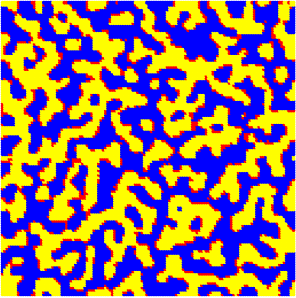}
}
\put(90,90)
{
 \includegraphics[width=.2\textwidth,height=.2\textwidth]{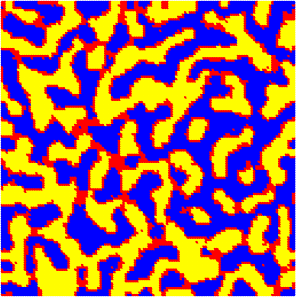}
}
\put(180,90)
{
 \includegraphics[width=.2\textwidth,height=.2\textwidth]{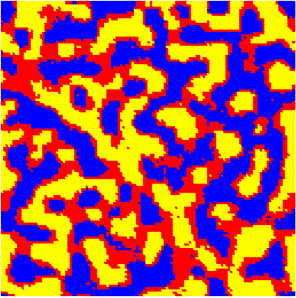}
}
\put(270,90)
{
 \includegraphics[width=.2\textwidth,height=.2\textwidth]{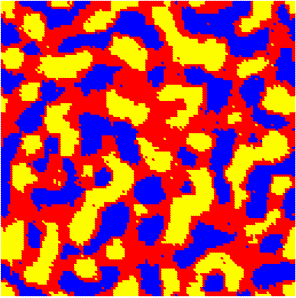}
}
\end{picture}
\caption{Configurations of the 3D Blume--Capel model 
with $J=1$ and $\beta=1.0$ observed on the $z$ constant section $(x,y,40)$, 
with $x,y=0,\dots,L-1$, of 
the $128^3$ 
lattice for 
$c_0=0.1,0.2,0.3,0.4,0.5,0.6,0.7,0.8$ (lexicographical order).
All the pictures refer to the iteration $2.0\times10^3$. As before
yellow, red, and blue points represent, respectively, minus, zero, 
and plus spins.
}
\label{fig:sto030}
\end{figure}

Indeed, as illustrated in Fig.~\ref{fig:sto010} at low solvent content, the dynamics 
tends to form interpenetrated structures of minuses and pluses separated by a thin 
layer of zeroes. In other words, the two polymers segregate into structures 
invading the whole lattice and are separated by an intermixed layer of solvent. 

The top picture in Fig. \ref{fig:sto010} is a 3D representation of three faces of the 
cubic lattice $\Lambda=\{0,\dots,L-1\}^3$: the bottom--left vertex is the 
origin $(0,0,0)$ and the two visible axes are the $x$ and the vertical $z$ axis.
Thus, as mentioned in the caption of the figure, 
the bottom--left, bottom--right, and top faces are, respectively, the sections $(x,0,z)$, $(L-1,y,z)$, and $(x,y,L-1)$ of the cube $\Lambda$, with 
$x,y,z=0,\dots,L-1$. In the bottom rows of the figure, the configuration of the system after 
$4.8\times10^3$ iteration is reported on six different square sections orthogonal to the $z$ 
axis. The different sections have been taken for values of the coordinate $z$ close to 
each other, so that one can get a clear idea of how these structures develop 
vertically, from the bottom towards the top, through the cubic lattice.

The fact that the walls of the formed domains 
are made of rather flat segments, 
giving the impression of a 3D labyrinth, is due
the rather low value of the temperature ($\beta=1.0$) which, given the 
high values of the energy differences involved in on single 3D spin swap, 
makes it 
improbable for a 
plus or a minus to abandon one cluster. 
The morphology appears completely different when the solvent content $c_0$ is increased. 
As reported in Fig.~\ref{fig:sto020}, the typical configuration is a mixtures 
of the three components with clusters of pluses and minuses surrounded by zeroes. 
We do not enter into a detailed description of the different part of the figures since they are analogous to those of Fig.~\ref{fig:sto010}.
The observed morphology is reminiscent of the ball--like structure 
that is observed in 2D. 
This is a relevant, and to some extent not completely expected fact, 
since 
in 3D the probability for two different 
spins to meet during their essentially random walk in the solvent background, 
so that a cluster can be formed, is small. We note that we were forced to use 
a small value of the temperature ($\beta=1.0$), to avoid cluster disintegration.
In other words, in this regime, the possibility to observe a cluster depends on the balance
between the time needed by two homologous 
spins to meet and that needed by a spin belonging 
to a cluster to detach. It is very interesting to note that this kind of 
behavior is similar to what happens in the metastable regime, where the typical time needed to form a critical droplet and to perform the transition to the stable state, is a consequence of the 
balance \cite{CO1996,cjs2024,cjs2025} between the contraction (detaching of a spin from a cluster) and the growing 
(absorption of a new particle by a cluster) times.

\begin{figure}[!t]
\begin{picture}(450,150)(-8,0)
\put(0,0)
{
 \includegraphics[width=.17\textwidth,height=.17\textwidth]{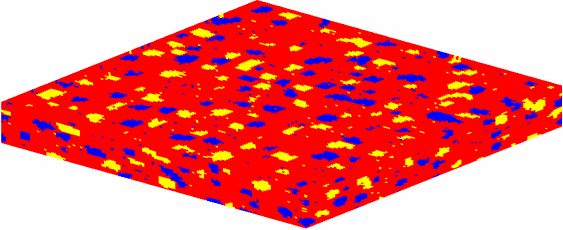}
}
\put(75,0)
{
 \includegraphics[width=.17\textwidth,height=.17\textwidth]{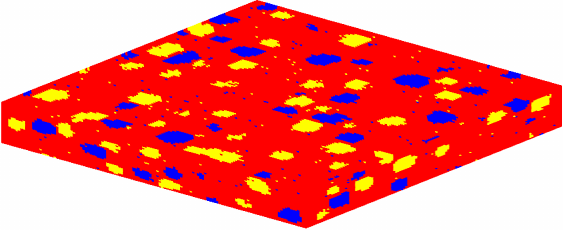}
}
\put(150,0)
{
 \includegraphics[width=.17\textwidth,height=.17\textwidth]{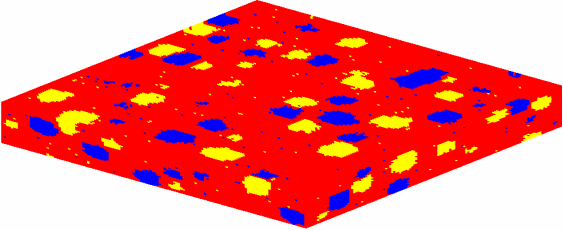}
}
\put(225,0)
{
 \includegraphics[width=.17\textwidth,height=.17\textwidth]{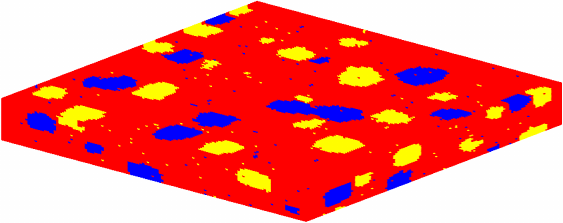}
}
\put(300,0)
{
 \includegraphics[width=.17\textwidth,height=.17\textwidth]{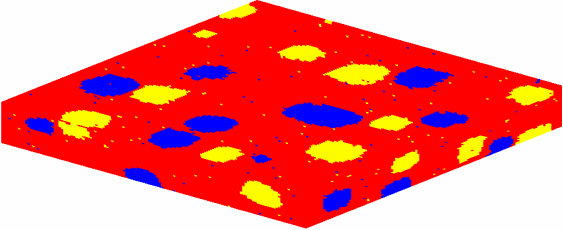}
}
\put(0,75)
{
 \includegraphics[width=.17\textwidth,height=.17\textwidth]{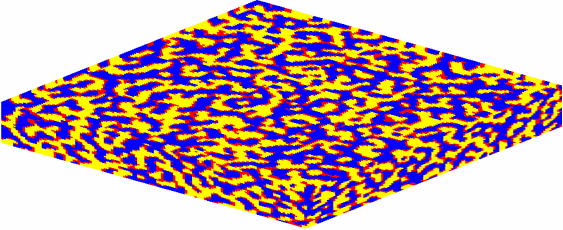}
}
\put(75,75)
{
 \includegraphics[width=.17\textwidth,height=.17\textwidth]{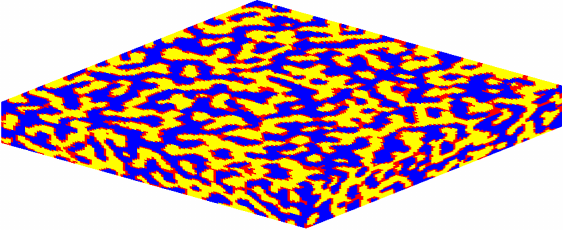}
}
\put(150,75)
{
 \includegraphics[width=.17\textwidth,height=.17\textwidth]{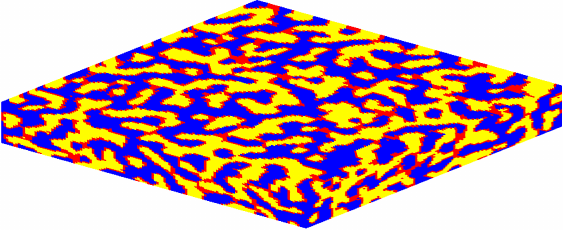}
}
\put(225,75)
{
 \includegraphics[width=.17\textwidth,height=.17\textwidth]{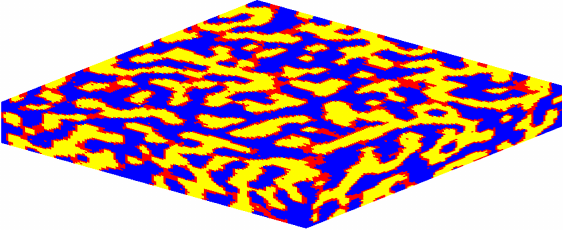}
}
\put(300,75)
{
 \includegraphics[width=.17\textwidth,height=.17\textwidth]{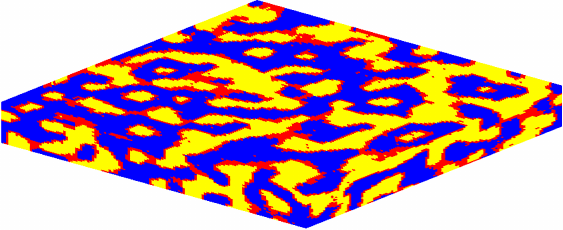}
}
\end{picture}
\caption{Configurations of the 3D Blume--Capel model 
with $J=1$ and $\beta=1.0$ observed on slabs of height 32 
of the $128^3$ 
lattice for 
$c_0=0.2$ (top row) and $c_0=0.8$ (bottom row). 
From the left to the right the configurations refer to 
iterations $200,800,1400,2400,4400$.
In all the pictures 
the left, right, and top planes are, respectively, the faces 
$(x,0,z)$, 
$(31,y,z)$, 
and $(x,y,31)$, 
with $x,y,z=0,\dots,127$
and
$z=0,\dots,31$.
}
\label{fig:sto040}
\end{figure}

In Fig.~\ref{fig:sto030},  we show the 
configuration of the system on a section of the cube orthogonal to the $z$ axis after 
the number of iterations specified in the caption. The different pictures 
refer to several values of the solvent concentration $c_0$ ranging 
from $0.1$ to $0.8$.
The figure show neatly how, with increasing the value of the solvent content, the observed morphology switches from the labyrinthic to the separated 
balls one. 

\begin{figure}[!t]
\begin{picture}(450,100)(-12,0)
\put(0,0)
{
 \includegraphics[width=.25\textwidth,height=.25\textwidth]{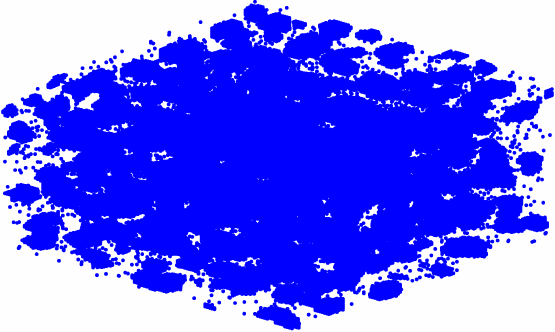}
}
\put(120,0)
{
 \includegraphics[width=.25\textwidth,height=.25\textwidth]{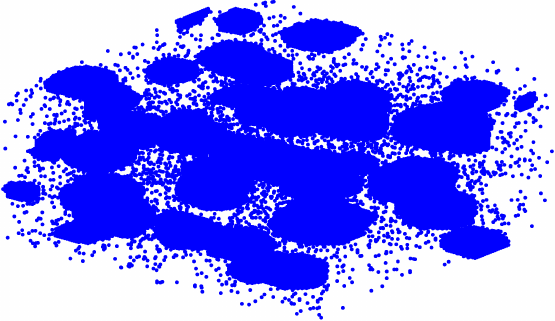}
}
\put(240,0)
{
 \includegraphics[width=.25\textwidth,height=.25\textwidth]{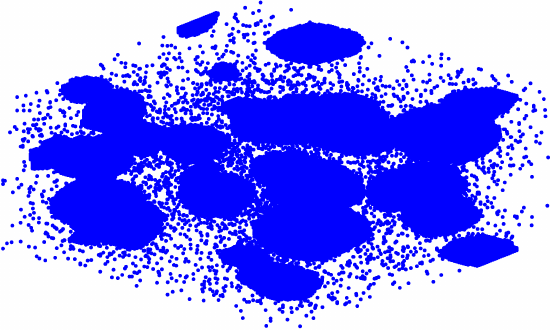}
}
\end{picture}
\caption{Plus configurations of the 3D Blume--Capel model 
with $J=1$ and $\beta=1.0$ observed on 
the $128^3$ 
lattice for 
$c_0=0.8$. 
From the left to the right the configurations refer to 
iterations $800,10800,17800$.
}
\label{fig:sto050}
\end{figure}

Finally, we comment briefly on Fig.~\ref{fig:sto040}, where we show the 
configuration of the system on the boundary of a portion of the lattice 
with base $128\times128$ and height $32$ (for the details see the caption 
of the figure). From the left to the right we plot the state 
after $200$, $800$, $1400$, $2400$, and $4400$ iterations. We recall that in one iteration $3L^3$ bonds, that is to say, 
$3\times128^3=6.291.456$ bonds in the case of the figure, are updated. 
The image gives a precise idea of the involved time scales and how 
they depend on the solvent concentration. Indeed, in the low solvent case, 
namely, $c_0=0.2$, after $200$ iterations the initial random 
configuration has been replaced by the labyrinth made of 
interpenetrated plus and minus structures, which are well formed. 
Its coarsening is started, and very slowly progresses in time as shown in 
the remaining pictures taken at larger times. Similarly, for $c_0=0.8$, 
at time $200$ very small domains start to appear. The following images 
show that the growing process progresses slowly, but constantly in time. 
The mechanism
is essentially made of two steps: evaporation 
of pluses and minus from small clusters followed by their capture
by the larger ones. 
A different representation of this process is provided in 
Fig.~\ref{fig:sto050}, where we have plotted in blue (as usual) the sole sites 
occupied by the pluses in the whole 3D lattice. The pictures from 
the left to the rigth report the configuration after $800$, $10800$, and 
$17800$, giving a quite detailed description of the growing process. 

\section{Morphology formation in 3D via a continuum model}\label{continuum_model}

The spin lattice model in Section~\ref{lattice_model} offers key advantages, such as flexibility and computational efficiency.  As anticipated already in \cite{Euromech}, our interest is to investigate later the effect that solvent evaporation can have on 3D morphologies. However, we foresee problems when implementing efficiently evaporation mechanisms in lattice models. Interestingly, this issue can be dealt with in continuum models. With this in mind,  we seek a computational approach that can handle evaporation by transitioning from a lattice to a continuum model. This requires passing to the hydrodynamic limit in a suitably scaled Blume–Capel model. However, the passage to this limit is not technically feasible with the nearest-neighbor interactions used in Section~\ref{lattice_model}; instead, a modified version incorporating a Kac-type long-range interaction must be considered.

\subsection{Description of the continuum model}
We recall now the structure of the continuum model derived in 
\cite{Marra}. Essentially, for  a given $\gamma>0$,  one introduces the Hamiltonian $H_\gamma : \mathcal{X}\to \mathbb{R}$,
given by
\begin{equation}
\label{kac000}
H_\gamma (\sigma) 
= 
\frac{1}{2}\sum_{\substack{i \neq j \in \Lambda}}
J_\gamma (i-j)[\sigma(i)-\sigma(j)]^2 
-\lambda\sum_{i\in \Lambda} [\sigma(i)]^2 
-h\sum_{x\in \Lambda}\sigma(i), 
\end{equation}
for all $\sigma\in \mathcal{X}$. In this context, the function $J_\gamma: \mathbb{R}^3\to \mathbb{R}$ 
is referred to as the Kac potential function, i.e.,
\begin{equation}
\label{geigamma}
J_\gamma(r)=\gamma^3 J(\gamma r)
\end{equation}
for all $r\in \mathbb{R}^3$. In \eqref{geigamma}, it holds
$J\in C^2(\mathbb{R}^3)$ is such that $J(r)=J(-r)$ (symmetry), 
$\int_{\mathbb{R}^3} J(r)\mathrm{d}r=1$ (normalized to $1$), $J(r)=0$ 
if $|r|>1$ (supported in the unit ball). 
The parameter $\gamma^{-1}$ delimitates the range of the interaction.

For the Kac version of the model, an exact calculation of the free energy in the so called hydrodynamics 
limit $\gamma\to0$ was done in \cite{Marra}, relying strongly on  
the Lebowitz--Penrose approach introduced in \cite{lebowitzpenrose1966}.
Consequently, we are now exploring a system  of two coupled non-local drift-diffusion equations with solution $(m,\phi)$ satisfying 
\begin{eqnarray}\label{Eq:MainModel_withoutEvap}
    \partial_t m &=& \nabla \cdot \left[\nabla m - 2 \beta (\phi -m^2 ) (\nabla J * m) \right], \quad (t,x) \in (0,T)\times\Omega,\\
    \partial_t \phi &=& \nabla \cdot \left[ \nabla \phi - 2 \beta m (1 - \phi) (\nabla J * m) \right].
\end{eqnarray}
In \eqref{Eq:MainModel_withoutEvap}, $t$ and $x$ represent the time and space variable, while  $m=m(t,x)$ is the 
average spin density (also called \emph{magnetization}), and $\phi=\phi(t,x)$ represents the average 
squared spin density and $1-\phi(t, x)$ represents the \emph{solvent volume concentration}. 
Similarly to what we did in Section~\ref{lattice_model}, we let $\Omega \, \subset \mathbb{R}^3$ be a cube with spatially periodic boundary conditions.  We call  $T>0$ the final time of the process. At this stage, the value for $T$ is chosen arbitrarily. Later on, after we will extend this model with the possibility of the solvent to evaporate, the meaning of $T$ will be linked to the amount of evaporated solvent.  

As initial data, we take
\begin{equation}\label{inic}
m(t=0)=m_0  \mbox{ and } \phi(t=0)=\phi_0 \mbox{ in } \bar\Omega.
\end{equation}
Recalling \cite{Marra}, if we set $u := (m,\phi)$, our system  \eqref{Eq:MainModel_withoutEvap} admits a natural gradient flow structure. In other words, we can write 
\begin{equation}\label{Eq:GradStruct}
    \partial_t u = \nabla \cdot \left( M\nabla \frac{\delta \mathcal{F}}{\delta u}\right).
\end{equation}
The mobility matrix is given by  
\[M = \beta (1-\phi) \begin{bmatrix}
    \phi + \frac{\phi^2 - m^2}{1-\phi} & m \\
    m & \phi
\end{bmatrix},\]
while the free energy functional $\mathcal{F}$ takes the form
\[\mathcal{F}(u) = \int_\Omega f(u(t,x)) \mathrm{d}x 
+ \frac{1}{2} \int_\Omega \int_\Omega J(x-x') [m(t,x) - m(t,x')]^2 \mathrm{d}x' \mathrm{d}x. \]
Here, we have 
$f(u) = \phi - m^2
+ \beta^{-1} [\frac{1}{2}(\phi+m)\log(\phi+m) + \frac{1}{2}(\phi-m)\log(\phi-m)+ (1-\phi) \log(1-\phi) - \phi \log(2)]$. 
This free energy structure resembles remotely the Flory–Huggins free energy density typically used for polymeric solutions, which admits generalizations for a $N$-component mixture of polymers.  

In \cite{Marra}, $m$ is referred to as magnetization and $\phi$ as concentration. 
The relevant physical quantities in our context is $1-\phi$ represents the solvent fraction and $m$ is interpreted here as the combined density of the components in the mixture. 
Interestingly, the precise physical meaning is best understood at the level of the stochastic dynamics. 
Recalling the notation of Section \ref{lattice_model}, If the spin variable at the site  $i\in\Lambda \subset \mathbb{Z}^3$ is $\sigma(i)$, then under suitable scalings the empirical measures $\sum_{i\in \Lambda}\sigma(i)$ and $\sum_{i\in\Lambda}\sigma^2(i)$ recover in the (many-particle) limit  the densities $m$ and $\phi$, respectively. These limit objects are interpreted here as follows: For a subset $\Omega'\subset \Omega$, the quantity $\int_{\Omega'} m(t,x)\mathrm{d}x$ is the net spin in the set $\Omega'$, while $\int_{\Omega'}(1-\phi(t,x))\mathrm{d}x$ represents
the solvent fraction in $\Omega'$. 
The energy structure $f(u)$ and the interpretation as spin density also hint at the natural inequality $|m| \leq \phi \leq 1$ which can be shown to hold even in the presence of evaporation of the solvent \cite[Theorem 1.1]{lyons2024phase}.

\subsection{Simulations}
To handle the numerical simulation of the system \eqref{Eq:MainModel_withoutEvap}, endowed with the initial condition \eqref{inic} and with periodic boundary conditions, we use a standard finite volume scheme. A presentation of our scheme can be found in \cite{lyons2023continuum}. At this point, we refer the reader as well to \cite{CRAS_Rainey} where we discussed alternatives of the proposed finite volume scheme that are proven to dissipate correctly a suitable quasi-energy to be used in the context of the structure \eqref{Eq:GradStruct}.

\begin{figure}[htbp]
    \centering
    \begin{tikzpicture}
        \node[anchor=south west, inner sep=0] (image) at (0.5, 0){\includegraphics[width=12cm, height=8cm]{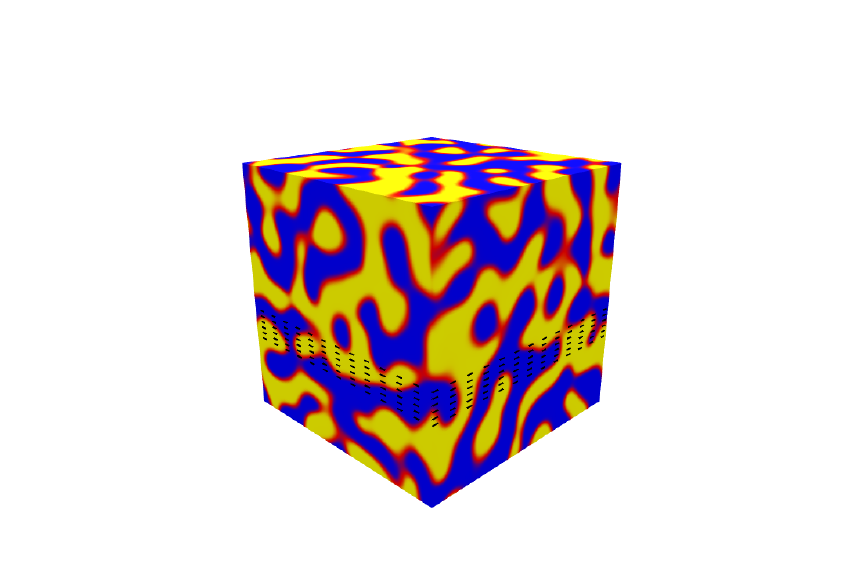}};
        \node[anchor=south west, inner sep=0] (image) at (10,1) 
            {\includegraphics[width=5cm, height=0.3cm, angle=90]{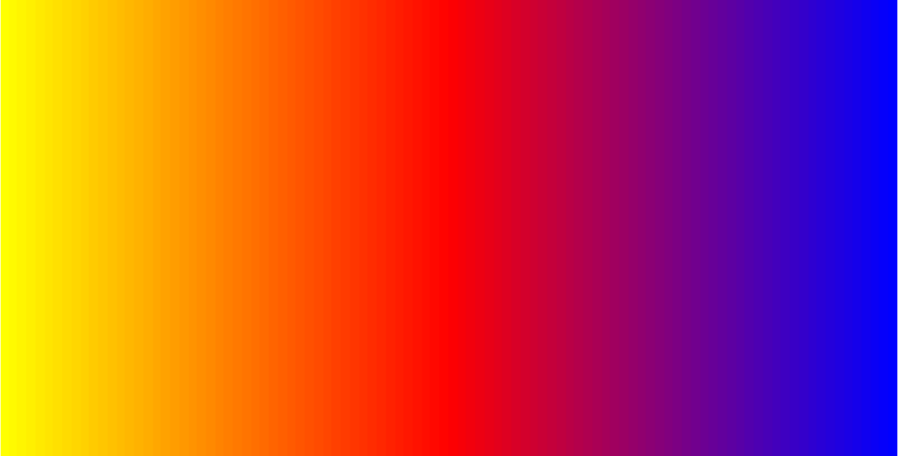}};
            \draw[thin, black, -] (10.15,1) -- (10.45, 1);
            \draw[thin, black, -] (10.15,3.5) -- (10.45,3.5);
            \draw[thin, black, -] (10.15,6.0) -- (10.45,6.0);
            \node at (10.8, 1.05) {-1};
            \node at (10.8, 3.55) {0};
            \node at (10.8, 6.05) {1};
    \end{tikzpicture}
    \\
    \subfloat{\includegraphics[width=0.3\textwidth]{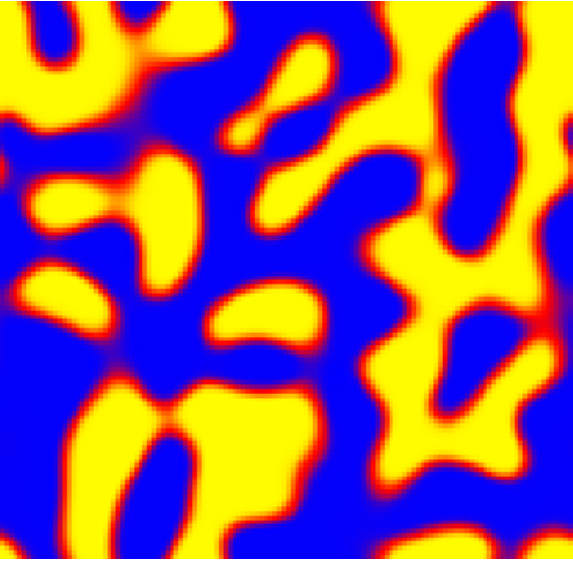}}\hspace{0.5cm}
    \subfloat{\includegraphics[width=0.3\textwidth]{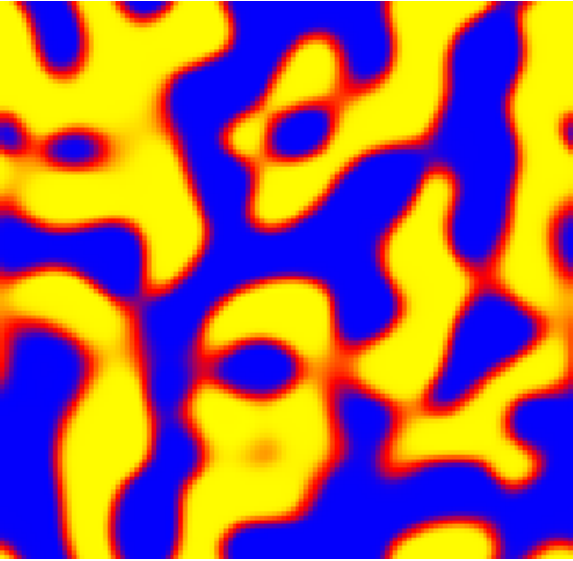}}\hspace{0.5cm} 
    \subfloat{\includegraphics[width=0.3\textwidth]{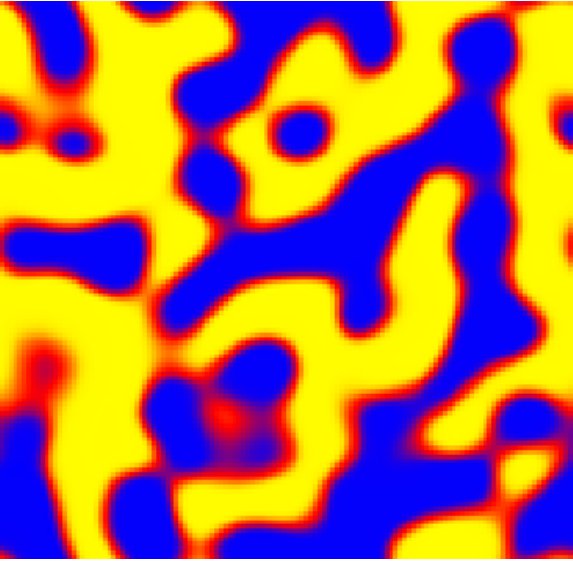}}\hspace{0.5cm} \\
    \subfloat{\includegraphics[width=0.3\textwidth]{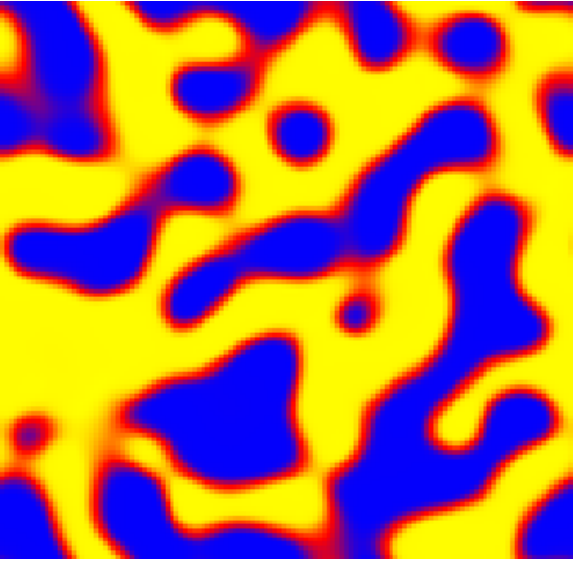}}\hspace{0.5cm}
    \subfloat{\includegraphics[width=0.3\textwidth]{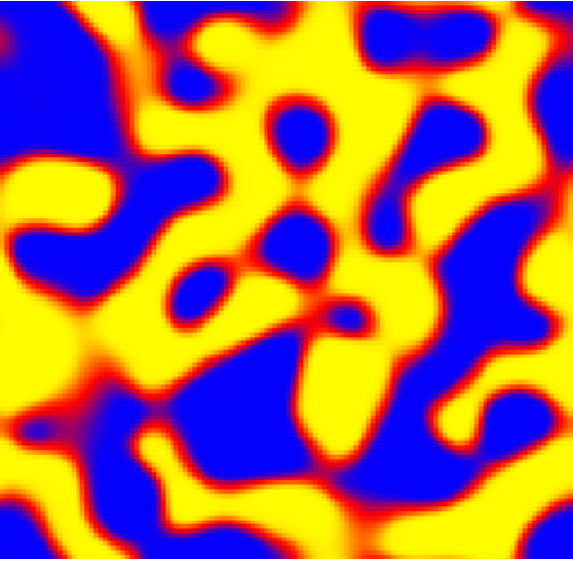}}\hspace{0.5cm}
    \subfloat{\includegraphics[width=0.3\textwidth]{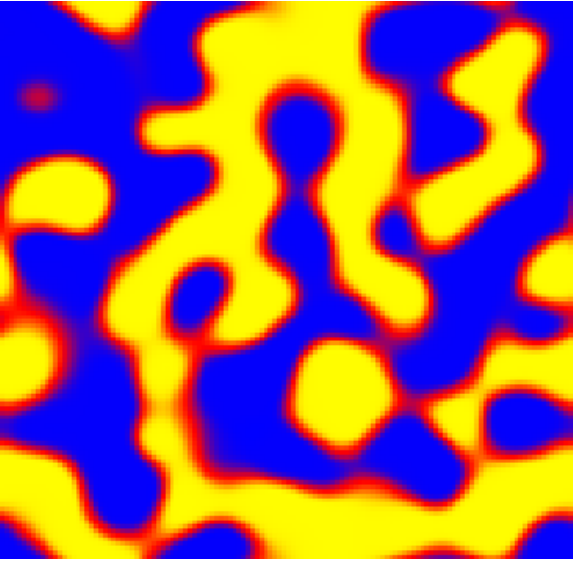}}\hspace{0.5cm}\\
    \caption{The cube on top shows $m(t, x)$ on the boundary of the computational domain, in the setting where $\phi_0 = 0.8$. This represents low solvent content. The cube is marked with six dashed line, cutting the cube horizontally. The dashed lines in ascending order are read in the heatmaps, left to right, up down. The bottom left corner of the heatmaps is the corner of the cube facing the reader.}
    \label{fig:cubeM_02}
\end{figure}

\begin{figure}[htbp]
    \centering
    \begin{tikzpicture}
        \node[anchor=south west, inner sep=0] (image) at (0.5, 0){\includegraphics[width=12cm, height=8cm]{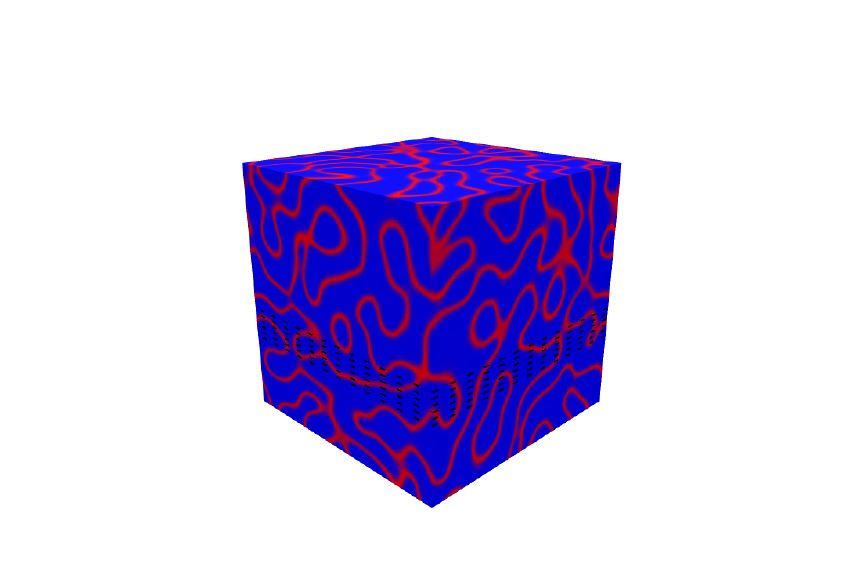}};
        \node[anchor=south west, inner sep=0] (image) at (10,1) 
            {\includegraphics[width=5cm, height=0.3cm, angle=90]{Figures/ContinuumFigs/colorbar.pdf}};
            \draw[thin, black, -] (10.15,1) -- (10.45, 1);
            \draw[thin, black, -] (10.15,3.5) -- (10.45,3.5);
            \draw[thin, black, -] (10.15,6.0) -- (10.45,6.0);
            \node at (10.8, 1.05) {-1};
            \node at (10.8, 3.55) {0};
            \node at (10.8, 6.05) {1};
    \end{tikzpicture}
    \\
    \subfloat{\includegraphics[width=0.3\textwidth]{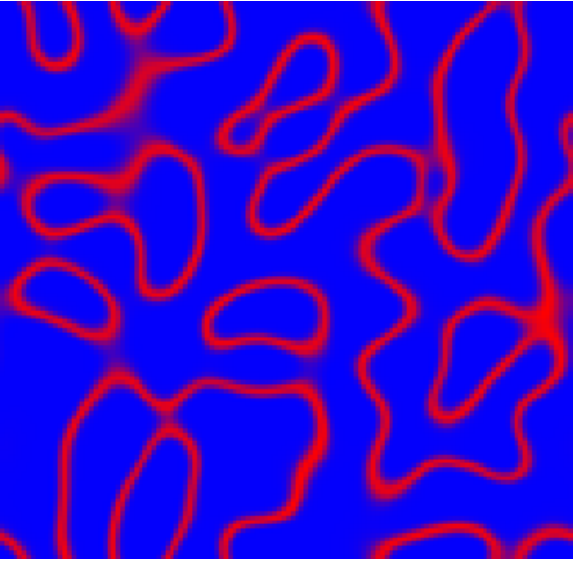}}\hspace{0.5cm}
    \subfloat{\includegraphics[width=0.3\textwidth]{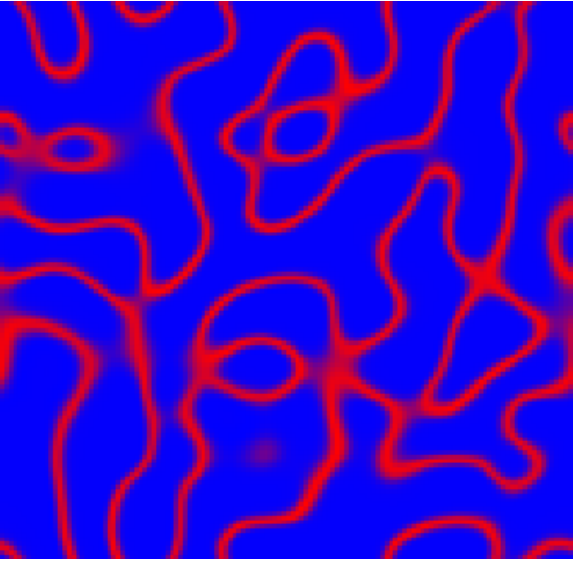}}\hspace{0.5cm} 
    \subfloat{\includegraphics[width=0.3\textwidth]{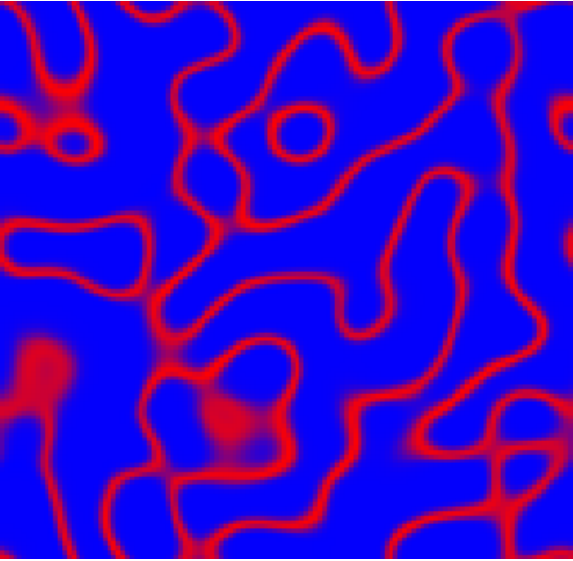}}\hspace{0.5cm} \\
    \subfloat{\includegraphics[width=0.3\textwidth]{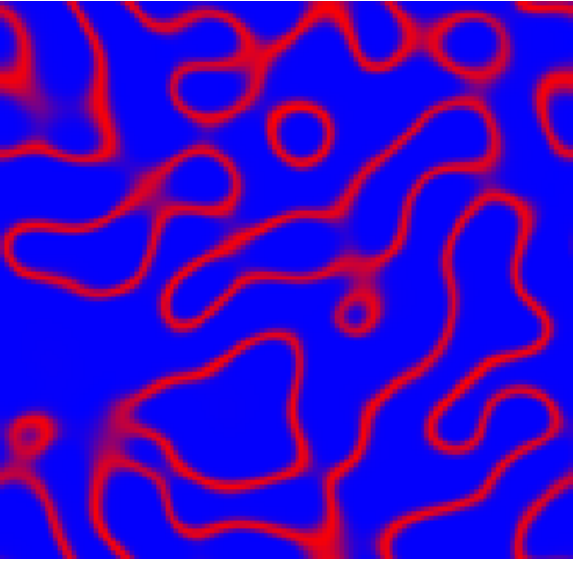}}\hspace{0.5cm}
    \subfloat{\includegraphics[width=0.3\textwidth]{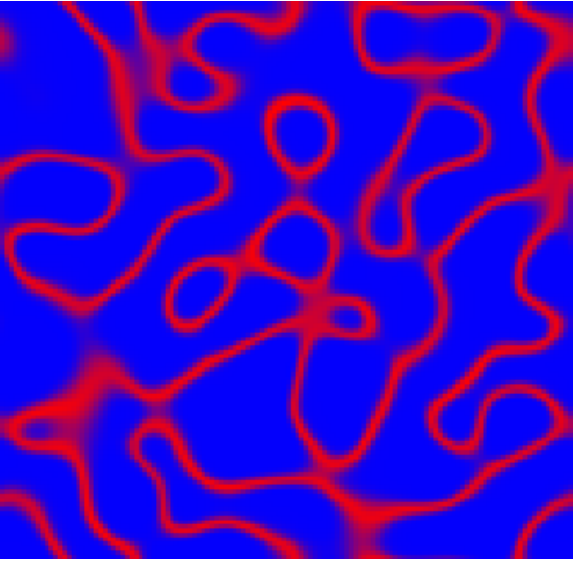}}\hspace{0.5cm}
    \subfloat{\includegraphics[width=0.3\textwidth]{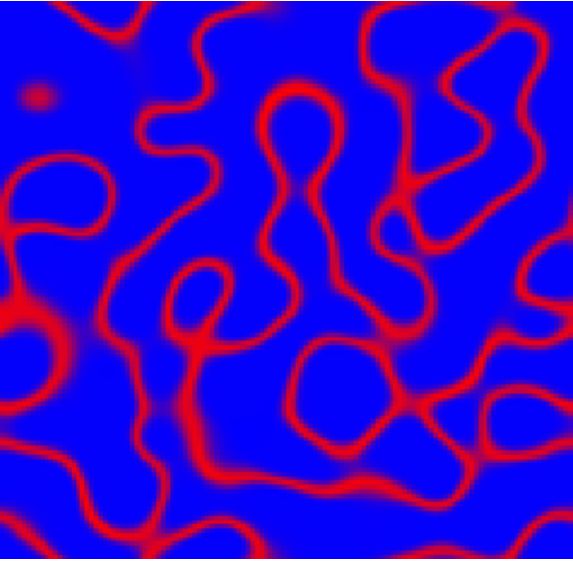}}\hspace{0.5cm}\\
    \caption{The cube on top shows $\phi(t, x)$ on the boundary of the computational domain, in the setting where $\phi_0 = 0.8$. This represents low solvent content. The cube is marked with six dashed line, cutting the cube horizontally. The dashed lines in ascending order are read in the heatmaps, left to right, up down. The bottom left corner of the heatmaps is the corner of the cube facing the reader.}
    \label{fig:cubeF_02}
\end{figure}

In our visualizations for $\phi$, we will make use of the color red to point out the spacial distribution of the solvent fraction ($\phi$ close to zero) and the color blue to indicate the presence of solute ($\phi$ close to one), independently of the polymer type. Moreover, if the values taken by the image of $m$ are close to $-1$, and respectively $+1$, then we point out the spacial distribution of the other two competing phases. We visualize these phases with yellow and
blue colors, respectively. Coherently with the color choice for the field $\phi$, 
the regions where  $m$ is near zero are colored red. Finally, the pink lines in the isosurface plots surrounds one of the components. The meaning and usage of the colors (where applicable) is in agreement also with the lattice-based simulation results shown in Section \ref{lattice_model:model}. 

The main goal of the simulations is investigating the effect of solvent concentration on the morphologies being formed. We find two distinct regimes for low and high solvent content. In particular, we will discuss the two cases $\phi_0 = 0.2$ (high solvent content) and $\phi_0=0.8$ (low solvent content), both with $\beta=1$ (recall that $\phi_0$ is the initial condition \eqref{inic}). 

\begin{figure}[htbp]
    \begin{tikzpicture}
        \node[anchor=south west, inner sep=0] (image) at (-3.5,0) 
            {\includegraphics[width=6.0cm]{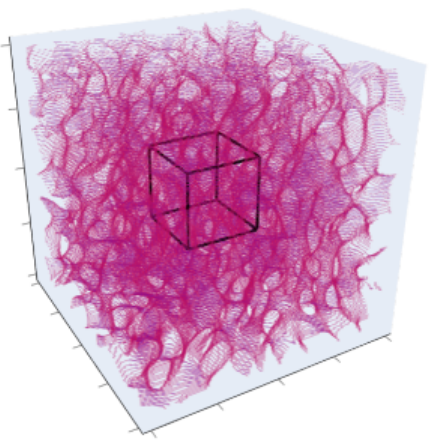}};
        \node[anchor=south west, inner sep=0] (image) at (3.5, 0){\includegraphics[width=6.0cm]{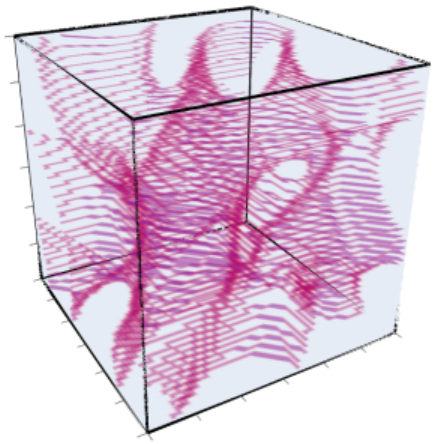}};
            \draw[thin, black, dashed] (-0.5,4.3) -- (7,6.05);
            \draw[thin, black, dashed] (-0.85,2.65) -- (5.5,0.2);
    \end{tikzpicture}
    \caption{In this figure we see the interface surface between regions dominated by one component and the solvent for a simulation with $\phi_0=0.8$. This represents low solvent content. The left image is the entire computational domain as well as a marked box, the content of which the right image shows.}
    \label{fig:contours02}
\end{figure}

In Fig.~\ref{fig:cubeM_02} and Fig.~\ref{fig:cubeF_02} we show 
the fields $m(t, x)$ and $\phi(t, x)$ at 
the boundaries of the computational domain and at some instances of the $x, y$-plane demarcated by dashed lines on the cube. 
In this regime, in which the solvent concentration is low, the morphology is characterized by interpenetrated domains containing high concentration of the two different polymer components. By visual inspection one is tempted to call this configuration bicontinuous.
Combining the information in Fig.~\ref{fig:cubeM_02} and Fig.~\ref{fig:cubeF_02} it is clear that the two components are separated from each other with a thin surface of solvent in between.
With this in mind we may use Fig.~\ref{fig:contours02} to get an idea of the three dimensional shapes we observe for this simulation.

\begin{figure}[htbp]
    \centering
    \begin{tikzpicture}
        \node[anchor=south west, inner sep=0] (image) at (0.5, 0){\includegraphics[width=12cm, height=8cm]{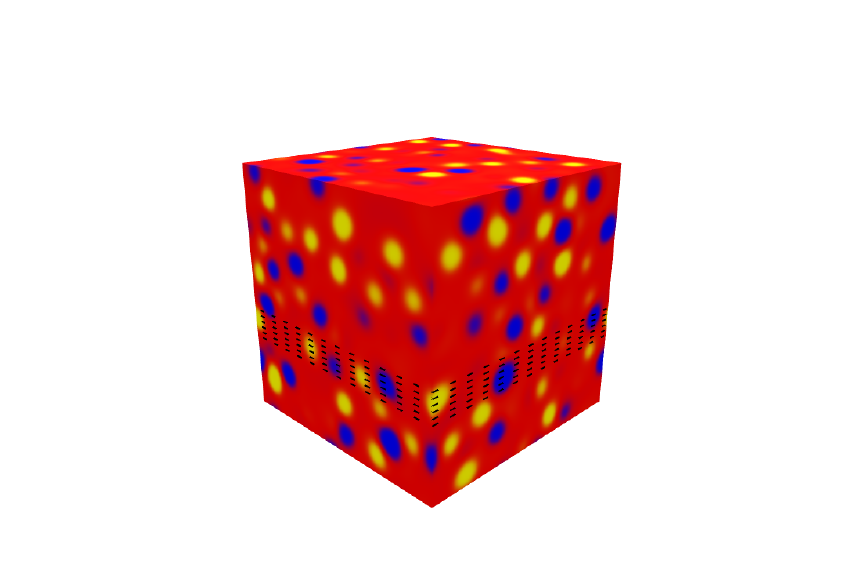}};
        \node[anchor=south west, inner sep=0] (image) at (10,1) 
            {\includegraphics[width=5cm, height=0.3cm, angle=90]{Figures/ContinuumFigs/colorbar.pdf}};
            \draw[thin, black, -] (10.15,1) -- (10.45, 1);
            \draw[thin, black, -] (10.15,3.5) -- (10.45,3.5);
            \draw[thin, black, -] (10.15,6.0) -- (10.45,6.0);
            \node at (10.8, 1.05) {-1};
            \node at (10.8, 3.55) {0};
            \node at (10.8, 6.05) {1};
    \end{tikzpicture}
    \\
    \subfloat{\includegraphics[width=0.3\textwidth]{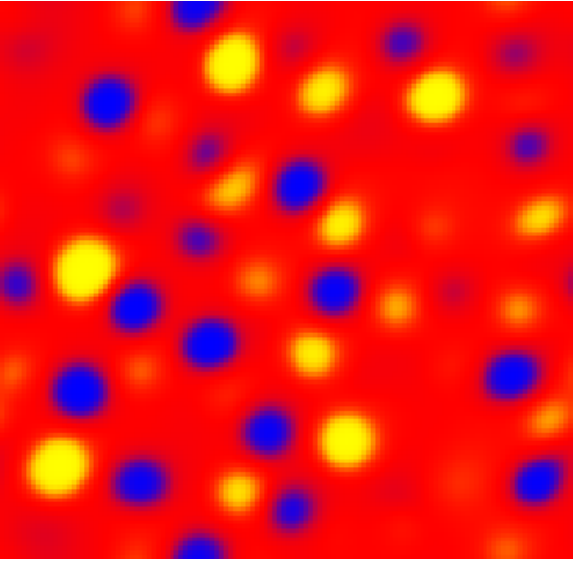}}\hspace{0.04cm}
    \subfloat{\includegraphics[width=0.3\textwidth]{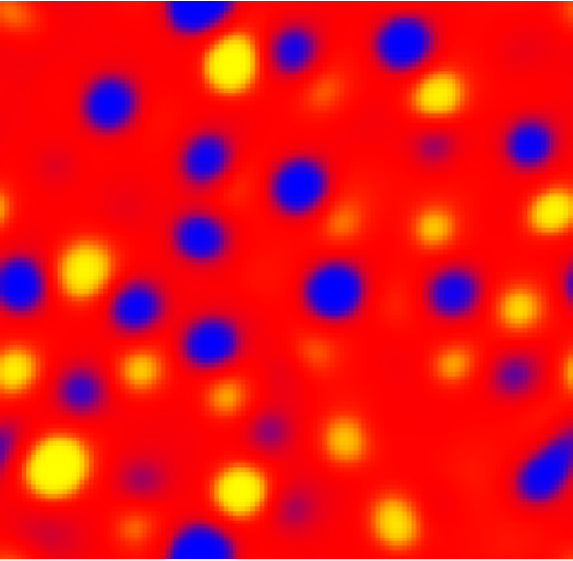}}\hspace{0.04cm} 
    \subfloat{\includegraphics[width=0.3\textwidth]{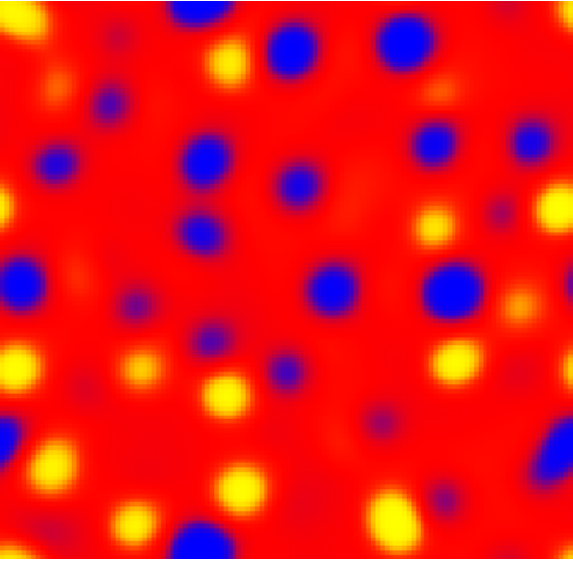}}\hspace{0.04cm} \\
    \subfloat{\includegraphics[width=0.3\textwidth]{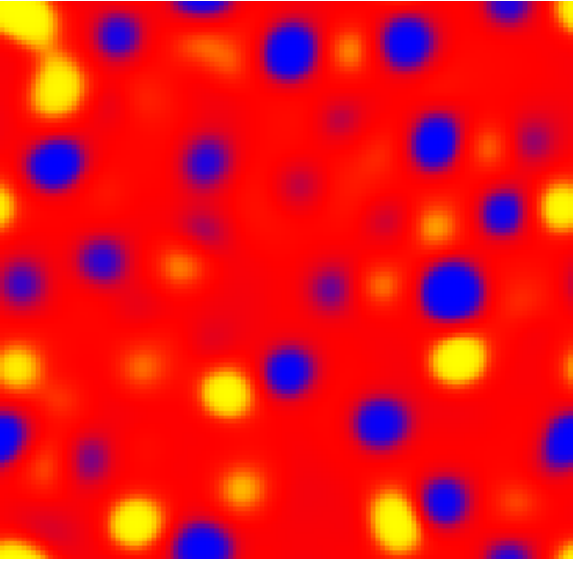}}\hspace{0.04cm}
    \subfloat{\includegraphics[width=0.3\textwidth]{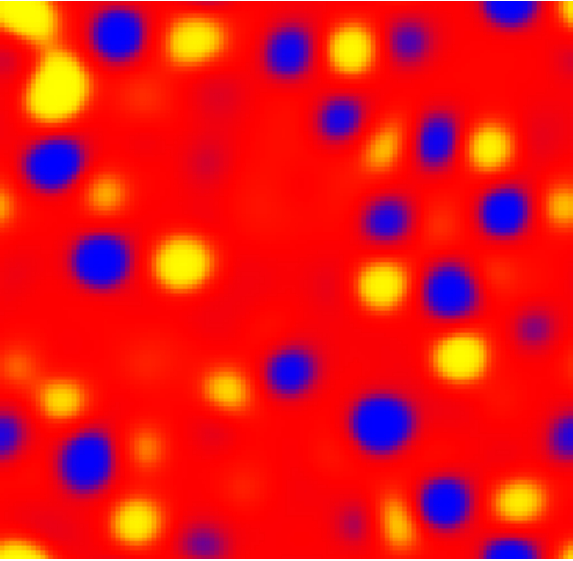}}\hspace{0.04cm}
    \subfloat{\includegraphics[width=0.3\textwidth]{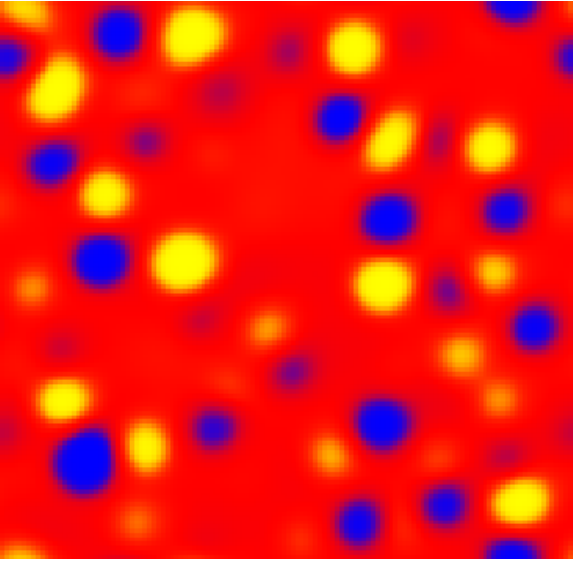}}\hspace{0.04cm}\\
    \caption{The cube on top shows $m(t, x)$ on the boundary of the computational domain, in the setting where $\phi_0 = 0.2$. This represents high solvent content. The cube is marked with six dashed line, cutting the cube horizontally. The dashed lines in ascending order are read in the heatmaps, left to right, up down. The bottom left corner of the heatmaps is the corner of the cube facing the reader.}
    \label{fig:cubeM_08}
\end{figure}

\begin{figure}[htbp]
    \centering
    \begin{tikzpicture}
        \node[anchor=south west, inner sep=0] (image) at (0.5, 0){\includegraphics[width=12cm, height=8cm]{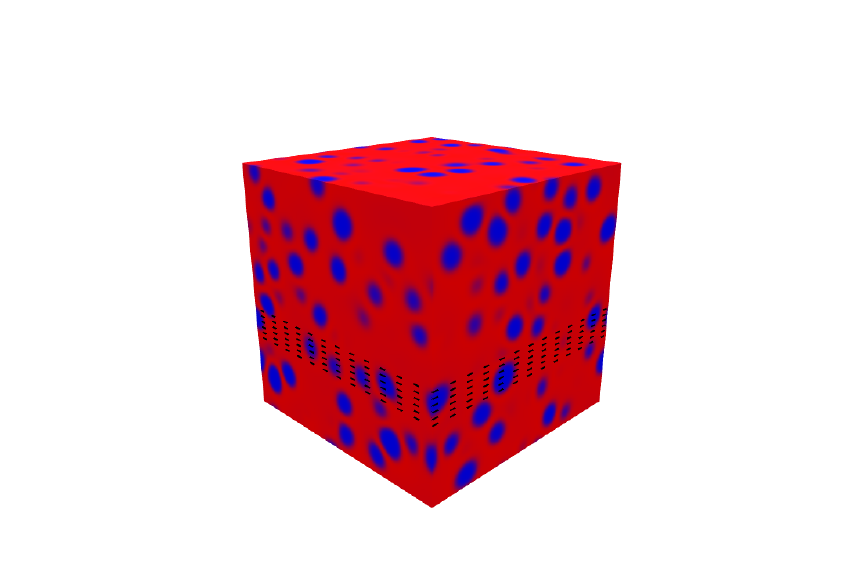}};
        \node[anchor=south west, inner sep=0] (image) at (10,1) 
            {\includegraphics[width=5cm, height=0.3cm, angle=90]{Figures/ContinuumFigs/colorbar.pdf}};
            \draw[thin, black, -] (10.15,1) -- (10.45, 1);
            \draw[thin, black, -] (10.15,3.5) -- (10.45,3.5);
            \draw[thin, black, -] (10.15,6.0) -- (10.45,6.0);
            \node at (10.8, 1.05) {-1};
            \node at (10.8, 3.55) {0};
            \node at (10.8, 6.05) {1};
    \end{tikzpicture}
    \\
    \subfloat{\includegraphics[width=0.3\textwidth]{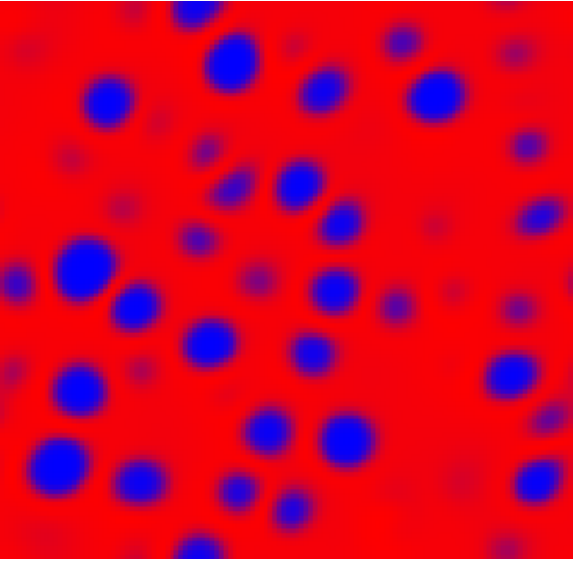}}\hspace{0.04cm}
    \subfloat{\includegraphics[width=0.3\textwidth]{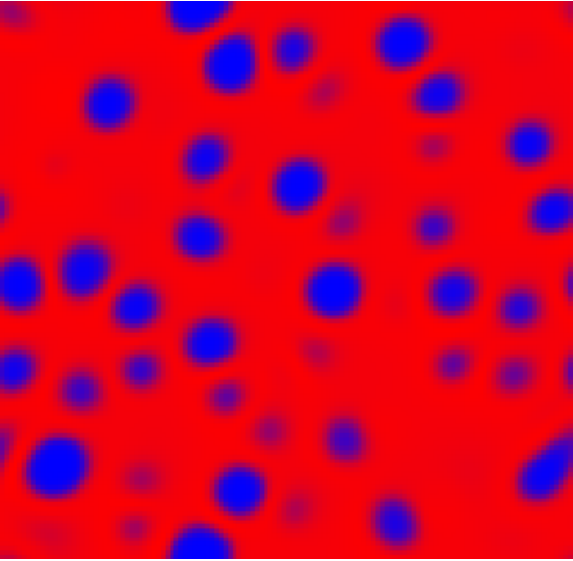}}\hspace{0.04cm} 
    \subfloat{\includegraphics[width=0.3\textwidth]{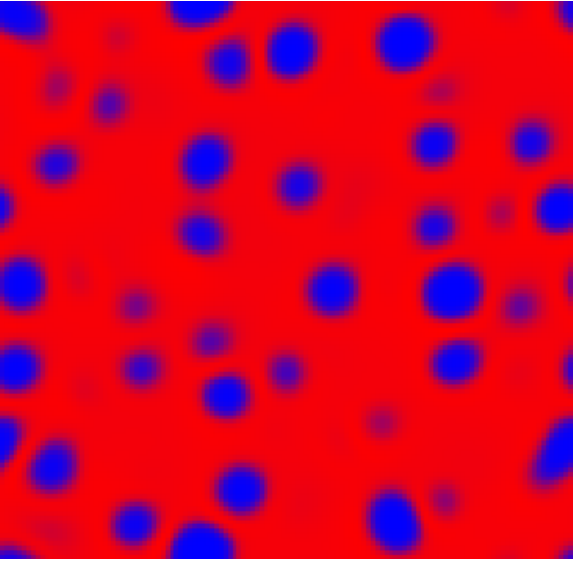}}\hspace{0.04cm} \\
    \subfloat{\includegraphics[width=0.3\textwidth]{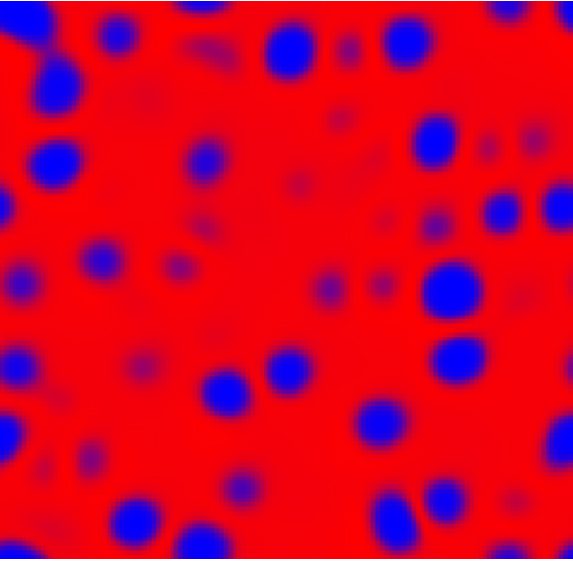}}\hspace{0.04cm}
    \subfloat{\includegraphics[width=0.3\textwidth]{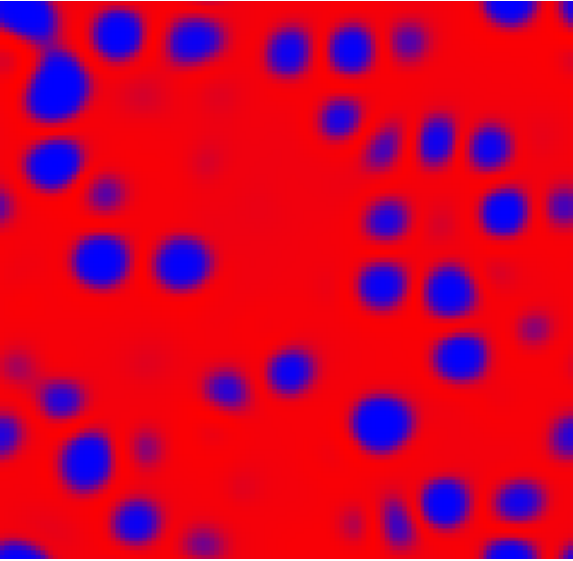}}\hspace{0.04cm}
    \subfloat{\includegraphics[width=0.3\textwidth]{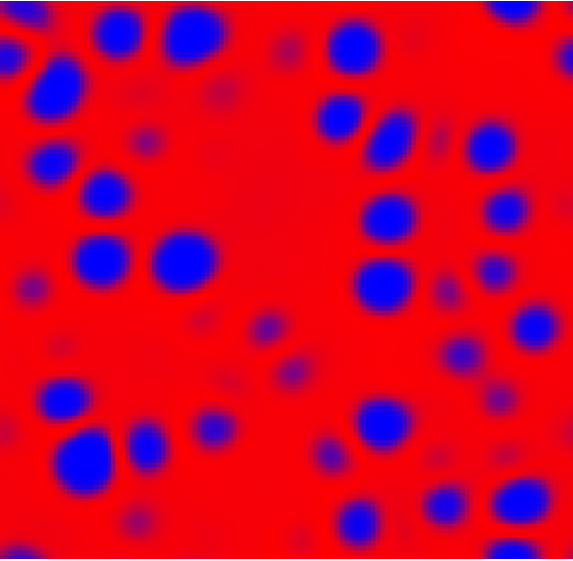}}\hspace{0.04cm}\\
    \caption{The cube on top shows $\phi(t, x)$ on the boundary of the computational domain, in the setting where $\phi_0 = 0.2$. This represents high solvent content. The cube is marked with six dashed line, cutting the cube horizontally. The dashed lines in ascending order are read in the heatmaps, left to right, up down. The bottom left corner of the heatmaps is the corner of the cube facing the reader.}
    \label{fig:cubeF_08}
\end{figure}

\begin{figure}[htbp]
    \begin{tikzpicture}
        \node[anchor=south west, inner sep=0] (image) at (-3.5,0) 
            {\includegraphics[width=6.0cm]{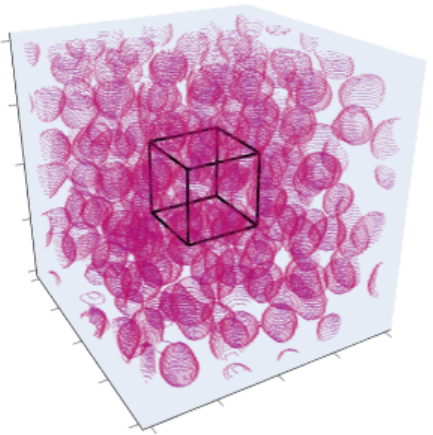}};
        \node[anchor=south west, inner sep=0] (image) at (3.5, 0){\includegraphics[width=6.0cm]{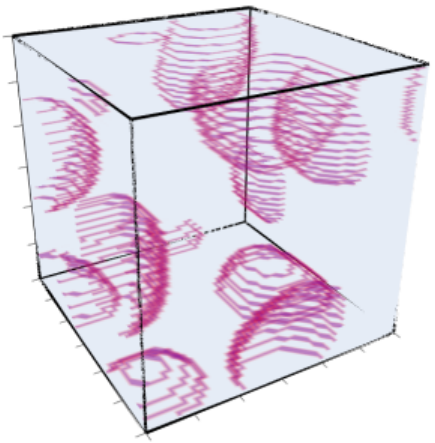}};
            \draw[thin, black, dashed] (-0.5,4.3) -- (7,6.05);
            \draw[thin, black, dashed] (-0.85,2.65) -- (5.5,0.2);
    \end{tikzpicture}
    \caption{In this figure we see the interface surface between regions dominated by one component and the solvent for a simulation with $\phi_0=0.2$. This represents high solvent content. The left image is the entire computational domain as well as a marked box, the content of which the right image shows.}
    \label{fig:contours08}
\end{figure}

In Fig.~\ref{fig:cubeM_08} and Fig.~\ref{fig:cubeF_08} we show 
the fields $m(t, x)$ and $\phi(t, x)$ at 
the boundaries of the computational domain and some representations of the $x, y$-plane whose position is demarcated by dashed lines on the cube. 
In this regime in which the solvent concentration is high, the morphology are dominated by isolated ball--like domains. This is clearly shown in the bottom rows of the two figures, 
in which mutually close by sections of the domain are shown. The images immediately suggest that these domains have small sizes in the $x$ and $y$ directions. But, looking at them in lexicographical order, that is to say shifting our gaze upward in the core of the domain, one  observes that these structures appear and disappear, which means that they resemble balls and not pillars, since they have limited extension also in the $z$ direction.
This observation is further underlined by looking at Fig.~\ref{fig:contours08}, where the three dimensional shapes of regions containing one component are shown. Indeed, in this high solvent concentration case, the pink lines isolate regions rich in  one particular component in the background sea of solvent.

\section{Conclusion}\label{conclusion}
In this study, we have demonstrated the ability to generate diverse morphology classes through numerical simulations in 3D by means of both lattice and continuum models, with a particular focus on structures suitable for organic solar cells (OSCs). Our results provide valuable insights into the design and optimization of thin-film morphologies, which play a crucial role in determining the efficiency and performance of OSCs. Additionally, the versatility of our approach suggests potential applications beyond OSCs, extending to other composite thin-film materials, such as adhesive bands.

Future work could explore further refinements of the simulation methods, incorporating additional physical components (interaction matrix, evaporation models, etc.) to enhance the predictive power of our models especially what concerns the transport of charges. Moreover, experimental validation of our simulated morphologies could help bridge the gap between theoretical modeling and practical implementation. By expanding the scope of our study, we aim to contribute to the broader field of nanostructured materials and their technological applications.

Finally, we highlight potential improvements to the numerical simulations that could enhance their efficiency. 
First, for the lattice model discussed in Section \ref{lattice_model}, clustering algorithms such as the Wolff \cite{wolff1989collective} and Swendsen–Wang \cite{swendsen1987nonuniversal} methods enable more efficient spin updates of non-conservative dynamics and have been successfully applied to 3D implementations of the Ising model.
Similar methods applicable to three-state conservative dynamics could enable simulations with larger lattice sizes and longer time scales. However, challenges may arise when accounting for the evaporation of the solvent component.
Next, for the finite volume scheme used in Section \ref{continuum_model}, it was shown in \cite{CRAS_Rainey} that due to the sharp interfaces formed on during the phase separation process, overshooting of the physical bounds of the solution can happen during simulations at coarse mesh sizes. 
This can be corrected using either flux-limiter techniques similar to those found in \cite{leveque1992numerical} or by bound-preserving gradient flow based schemes like those discussed in \cite{bailo2023,CRAS_Rainey}. 
These schemes have the added benefit of provably dissipating the energy structure \eqref{Eq:GradStruct} (though we point out that numerical experiments suggest standard finite volume schemes also dissipate the free energy).

\section*{Acknowledgments}
\noindent
NJ and AM are involved in Swedish Energy Agency's project Solar Electricity Research Centre (SOLVE) with grant number 52693-1. We acknowledge the support of KK-NU project 20230010-H-01 for facilitating  the visit of ENMC to Karlstad during  March 2025, when this manuscript has been completed. The computations were enabled by resources provided by the National Academic Infrastructure for Supercomputing in Sweden (NAISS), partially funded by the Swedish Research Council through grant agreement no. 2022-06725.
ENMC thanks the PRIN 2022 project
``Mathematical Modelling of Heterogeneous Systems (MMHS)",
financed by the European Union - Next Generation EU,
CUP B53D23009360006, Project Code 2022MKB7MM, PNRR M4.C2.1.1.
ENMC thanks, also, the Mathematics Department of the Karlstad University for warm hospitality
and GNFM--INDAM.




\end{document}